\documentclass[aps,showpacs,pra,twocolumn]{revtex4}
\usepackage{epsfig,amssymb,amscd}

\begin{document}
\bibliographystyle{prsty}

\title{6$^{th}$ Order Robust Gates for Quantum Control}
\author{D. Mc Hugh\footnote[2]{Email: dmchugh@thphys.may.ie}, J. Twamley}
\affiliation{Department of Mathematical Physics,\\
National University of Ireland Maynooth,\\ Maynooth, Co. Kildare, Ireland}
\received{\today}

\begin{abstract}
Composite pulse sequences designed for nuclear magnetic resonance experiments 
are currently being applied in many quantum information processing
technologies.
We present an analysis of a family of composite pulse sequences used to address
systematic pulse-length errors in the execution of quantum gates. It has been demonstrated by Cummins {\it et al.}
in \cite{JJones}
that for this family of composite pulse sequences, the fidelity of the resulting
unitary operation compared with the ideal unitary operation is 
$1-C\epsilon^6$, where $\epsilon$ is the fractional error in the
length of the pulse. We derive an exact expression for the 6$^{th}$ order
coefficient, $C$, and from this deduce conditions under which 
this 6$^{th}$ order dependance is observed. We also present new pulse
sequences which achieve the same fidelity. 
\end{abstract}
\vskip 0.1cm
\maketitle

Systematic errors are always present in any experimental set-up. 
They are best dealt with by stripping down the experiment and finding them
one by one. Even after doing this, there is still the possibility of not 
eliminating
them all. Quantum computers operate universally with single-qubit rotations and 
a controlled NOT gate between two qubits \cite{deutsch}. 
Recently, Cummins {\it et al.} \cite{JJones} have addressed the issue of 
systematic errors in single-qubit gates due to pulse-length and off-resonance
effects. Off-resonance errors result in a rotation around an 
axis tilted with respect to the desired rotation axis. Pulse-length errors result in a rotation through an angle 
which falls short of, or goes beyond, the desired angle of 
rotation, due to an error in the timing of the pulse. 
Composite pulse techniques have received much attention as a means to correct
systematic errors \cite{wimperis, JJones,JJones1,jones_pb}, with 
\cite{Chuang} a significant recent development. First developed in the NMR setting, 
these techniques can be very 
useful in reducing the effects of systematic errors in a wide range of
quantum computer proposals. 
They have been identified as such in numerous quantum information processing
technologies such as trapped-ion 
technologies \cite{schmidt,gulde,wineland}, in rare-earth doped crystal
technology \cite{molmer,wesenberg}, in superconducting technologies
\cite{martinis,mooij} and in solid state quantum information processing 
technologies \cite{kane,fisher}.
The BB1 composite pulse sequence, first developed by Wimperis 
\cite{wimperis}, deals with pulse-length errors in a remarkably efficient 
manner. It is shown in \cite {JJones} that the composite pulse sequence has a 
fidelity of $1-C\epsilon^6$ when compared to the exact qubit rotation. 
In this article we examine the origin of this $6^{th}$ order dependence on
the error in the fidelity of the gate, and deduce some constraints on the
possible composite pulse sequences one may use. We also suggest some other 
pulse sequences whose fidelities display this 6$^{th}$ order dependence. 

A general single-qubit rotation around an axis in the $XY$-plane of the Bloch
sphere can be written as,
\begin{eqnarray}
R(\theta,\alpha)&=&\exp\left(-i\frac{\theta}{2}(X\cos\alpha+Y\sin\alpha) \right)
\end{eqnarray}

\noindent
where $\theta$ is the angle through which the qubit is rotated and 
$\alpha$ is the angle of the axis in the $XY$ plane of the Bloch sphere, 
$\alpha=0$ being 
the $X$ axis. We now consider the effects of a systematic error, such as a pulse-length
type error. Using a superscript to denote pulse sequences suffering this type of error, where all rotation angles are altered by the fractional error $\epsilon$,  the above one single-qubit rotation becomes
\begin{eqnarray}
R^\epsilon(\theta,\alpha)&=&\exp\left(-i\frac{\theta(1+\epsilon)}{2}(X\cos\alpha+Y\sin\alpha)\right),
\end{eqnarray}

\noindent
where $\epsilon$ is the fractional error. In order to compare the error-prone 
unitary transformation ($V=R^\epsilon (\theta,\alpha)$) with the error-free unitary transformation ($U=R(\theta,\alpha)$),
the following fidelity definition is used,
\begin{eqnarray}
\cal F &=&\frac{|Tr(VU^{\dagger})|}{Tr(UU^{\dagger})}.
\end{eqnarray}

\noindent
The composite pulse sequence as presented by Wimperis takes the form,
\begin{eqnarray}
W^\epsilon(\phi_1,\phi_2)&=&R^\epsilon(\pi,\phi_1)\,R^\epsilon(2\pi,\phi_2)\,R^\epsilon(\pi,\phi_1),
\end{eqnarray}

\noindent
so that when $\epsilon=0$, $W$ is simply the identity. When this pulse sequence
is carried out before or after the desired single-qubit rotation, the fidelity 
of the resulting composite pulse sequence is $1-C\epsilon^6$. In fact, 
$W^\epsilon(\phi_1,\phi_2)$ may be placed at any point during a rotation around
a given axis, i.e.,
\begin{eqnarray*}
R^\epsilon(a\theta,\alpha)\leftrightarrow W^\epsilon(\phi_1,\phi_2)\leftrightarrow R^\epsilon((1-a)\theta,\alpha),
\end{eqnarray*}

\noindent
with $a\in[0,1]$, and the same fidelity is observed.

\noindent
In order to show where this comes from, we need to examine the definition of the fidelity. 
First we let,
\begin{eqnarray*}
A&=&X\cos\phi_1+Y\sin\phi_1\\
B&=&X\cos\phi_2+Y\sin\phi_2\\
C&=&X\cos\alpha+Y\sin\alpha
\end{eqnarray*}

\noindent
be three axes in the $XY$-plane of the Bloch sphere. $C$ is the axis around
which we wish to perform the rotation and $A$ and $B$ are the two axes in $W^\epsilon(\phi_1,\phi_2)$. The fidelity is,
\begin{eqnarray*}
\cal F&=& \frac{1}{2}|Tr(e^{-i\frac{\theta(1+\epsilon)}{2}C}W^\epsilon(\phi_1,\phi_2)e^{i\frac{\theta}{2}C})|\\
&=&\frac{1}{2}|Tr(e^{-i\frac{\theta}{2}C}e^{-i\frac{\epsilon\theta}{2}C}W^\epsilon(\phi_1,\phi_2)e^{i\frac{\theta}{2}C})|\\
&=&\frac{1}{2}|Tr(e^{-i\frac{\epsilon\theta}{4}C}W^\epsilon(\phi_1,\phi_2)e^{-i\frac{\epsilon\theta}{4}C})|.
\end{eqnarray*}

\noindent
The last step is due to the trace property of invariance under cyclic 
permutations. $W^\epsilon(\phi_1,\phi_2)$ can be simplified to, 
\begin{eqnarray*}
W^\epsilon(\phi_1,\phi_2)&=&e^{-i\frac{\pi(1+\epsilon)}{2}A}
e^{(-i(1+\epsilon)\pi B)}e^{-i\frac{\pi(1+\epsilon)}{2}A}\\
&=&e^{-i\frac{\epsilon\pi}{2}A}(-iA)e^{-i\epsilon\pi B}(-{\mathbb{I}})(-iA)
e^{-i\frac{\epsilon\pi}{2}A}\\
&=&e^{-i\frac{\epsilon\pi}{2}A}e^{-i\epsilon\pi ABA}
e^{-i\frac{\epsilon\pi}{2}A}
\end{eqnarray*}

\noindent
since $A^2=\mathbb{I}=B^2$. $ABA = X\cos(2\phi_1-\phi_2)+Y\sin(2\phi_1-\phi_2)$ 
is another axis in the $XY$-plane of the Bloch sphere. The fidelity 
now takes on a more symmetric look when written as, 
\begin{eqnarray*}
\cal F&=& \frac{1}{2}|Tr(e^{-i\frac{\epsilon\theta}{4}C}
e^{-i\frac{\epsilon\pi}{2}A}
e^{-i\epsilon\pi ABA}e^{-i\frac{\epsilon\pi}{2}A}e^{-i\frac{\epsilon\theta}{4}C})|\\
\end{eqnarray*}

\noindent
This form for the fidelity turns out to be the reason the BB1 sequence performs
so well. The symmetric Baker-Campbell-Hausdorff formula is stated and derived 
in \cite{iserles} as,
\begin{eqnarray}
e^{\frac{1}{2}tR}e^{tS}e^{\frac{1}{2}tR}&=&e^{sbch(t;R,S)}
\end{eqnarray}

\noindent
with,
\begin{eqnarray*}
sbch(t;R,S)&=&t(R+S)-\frac{1}{24}t^3[R+2S,[R,S]]+{\mathcal O}(t^5) 
\end{eqnarray*}

\noindent
If we define $Q=-i\theta C$, $R=-i\pi A$ and $S=-i\pi ABA$ to simplify the
notation, we 
can reduce the expression in the trace above by twice applying the 
symmetric BCH formula, yielding,
\begin{eqnarray*}
{\cal F}&=&\frac{1}{2}|Tr(e^{-i\frac{\epsilon\theta}{4}C}e^{-i\frac{\epsilon\pi}{2}A}e^{-i\epsilon\pi ABA}e^{-i\frac{\epsilon\pi}{2}A}e^{-i\frac{\epsilon\theta}{4}C})|\\
&=&\frac{1}{2}|Tr(e^{\frac{\epsilon}{4}Q}e^{\frac{\epsilon}{2}R}e^{\epsilon S}e^{\frac{\epsilon}{2}R}e^{\frac{\epsilon}{4}Q})|\\
&=&\frac{1}{2}|Tr(e^{\frac{\epsilon}{4}Q}e^{\frac{\epsilon}{2}(\frac{2}{\epsilon}sbch(\epsilon;R,S))}e^{\frac{\epsilon}{4}Q})|\\
&=&\frac{1}{2}|Tr(e^{sbch(\frac{\epsilon}{2};Q,\frac{2}{\epsilon}sbch(\epsilon;R,S))})|
\end{eqnarray*}
 
\noindent
Letting $P(\epsilon)=sbch(\frac{\epsilon}{2};Q,\frac{2}{\epsilon}sbch(\epsilon;R,S))$, we find that, 
\begin{eqnarray*}
P(\epsilon)&=&\epsilon(\frac{1}{2}Q+R+S)+\epsilon^3\left(-\frac{1}{24}[R+2S,[R,S
]]\right.\\&&\left.-\frac{1}{192}[Q,[Q,2(R+S)]-\frac{1}{24}[R+S,[Q,R+S]]\right)\\&&+{\cal O}(\epsilon^5).
\end{eqnarray*}

\noindent
Given that all the operators in $P(\epsilon)$ are proportional to the Pauli
operators, $Tr(P(\epsilon))=0$ and therefore,
\begin{eqnarray*}
Tr(e^{P(\epsilon)})&=&Tr(\mathbb{I}) + Tr(\frac{1}{2}(P(\epsilon))^2)\\
\Rightarrow{\cal F}&=&1+\frac{1}{4}Tr((P(\epsilon))^2)+\cdots
\end{eqnarray*}

For the next step we let the first derivative of the total pulse sequence with
respect to the error, $\epsilon$, equal the zero matrix at $\epsilon=0$ to 
find a relation between the rotation axis, $C$, and the axes 
$A$ and $B$,

The total pulse sequence is given by,
\begin{eqnarray*}
BB(\epsilon)&=&e^{\frac{(1+\epsilon)}{2}Q}e^{\frac{(1+\epsilon)}{2}R}e^{(1+\epsilon) S}e^{\frac{(1+\epsilon)}{2}R}
\end{eqnarray*}

\noindent
so that,
\begin{eqnarray*}
\left.\frac{dBB(\epsilon)}{d\epsilon}\right|_{\epsilon=0}&=&0\\
\end{eqnarray*}

$\Rightarrow\frac{1}{2}Q+(R+S)=0.$

\noindent
This allows us to simplify the expression for $P(\epsilon)$,
\begin{eqnarray*}
P(\epsilon)&=& \epsilon^3(-\frac{1}{24}[R+2S,[R,S]] + \frac{1}{192}[Q,[Q,Q]]\\&&-\frac{1}{96}[Q,[Q,Q]]) + {\cal O}(\epsilon^5)\\
&=&-\frac{1}{24}[R+2S,[R,S]]\epsilon^3+ {\cal O}(\epsilon^5).
\end{eqnarray*}
 
\noindent
Next we expand the commutator $[R+2S,[R,S]]$ to get,
\begin{eqnarray*}
([R+2S,[R,S]])^2&=&-\pi^6([A+2ABA,[A,ABA]])^2\\
&=&-\pi^6(40{\mathbb{I}}+8(AB+BA)\\&&-20(ABAB+BABA)\\&&-8(ABAB
AB+BABABA))\\
&=&-\pi^6[40+16\cos(\phi_2-\phi_1)
\\&&\ \ \ \ \ \ \ \ \ \ -\ 40\cos 2(\phi_2-\phi_1)
\\&&\ \ \ \ \ \ \ \ \ \ -\ 16\cos 3(\phi_2- \phi_1)]{\mathbb{I}}
\end{eqnarray*}

\noindent
since $AB= \exp(i(\phi_2-\phi_1)Z)$.

\noindent
Therefore, we can write the fidelity as,
\begin{eqnarray*}
{\cal F}&=&1+\frac{1}{2304}\epsilon^6Tr(([R+2S,[R,S]])^2)+{\cal O}(\epsilon^8)\\
&=&1-\frac{\epsilon^6\pi^6}{144}[5+2\cos(\phi_2-\phi_1)-5\cos 2(\phi_2-\phi_1) \\&&\ \ \ \ \ \ \ \ \ \ \ \  -2\cos 3(\phi_2-\phi_1)]+{\cal O}(\epsilon^8)
\end{eqnarray*}

\noindent
Finally, by writing $A$, $B$ and $C$ in terms of the Pauli operators, $X$ and $Y$,
the condition $\frac{1}{2}Q+R+S=0$ implies,
\begin{eqnarray}
\cos\phi_1+\cos(2\phi_1-\phi_2)&=&-\frac{\theta}{2\pi}\cos\alpha\\
\sin\phi_1+\sin(2\phi_1-\phi_2)&=&-\frac{\theta}{2\pi}\sin\alpha ,
\end{eqnarray}

\noindent
and these two equations lead to the following rule for choosing 
$\phi_1$ and $\phi_2$ given $\theta$ and $\alpha$,
\begin{eqnarray}
\phi_1&=&\alpha-\frac{1}{2}\arccos\left(\frac{\theta^2}{8\pi^2}-1\right)\\
\phi_2&=&3\phi_1-2\alpha
\end{eqnarray}

\vspace{6mm}

\noindent
To give a concrete example, we choose $\alpha=0$, $\theta=\pi$
corresponding to a $180^{\circ}$ rotation about the $X$ axis of the 
Bloch sphere. This gives us,
\begin{eqnarray*}
\phi_1&=&\arccos\left(-\frac{1}{4}\right)\\
\phi_2&=&3\phi_1\\
\end{eqnarray*}

\noindent
and a fidelity,
\begin{eqnarray*}
{\cal F}&=&1-\frac{5}{1024}\pi^6\epsilon^6 + {\cal O}(\epsilon^8).
\end{eqnarray*}

\noindent
which agrees with the findings in \cite{JJones1}.

It can be readily seen that if $n$ copies of $W^\epsilon(\phi_1,\phi_2)$ are 
carried out one after the other, then the condition that 
$\left.\frac{dBB}{d\epsilon}\right|_{\epsilon=0}=0$,
\begin{eqnarray*}
\Rightarrow \frac{1}{2}Q+n(R+S)=0
\end{eqnarray*}

\noindent
and also $P(\epsilon)=\epsilon(\frac{1}{2}Q+n(R+S))+{\cal O}(\epsilon^3)$. 
This defines the general $Wn$ pulses introduced in \cite{JJones} as,
\begin{eqnarray*}
\phi_1&=&\alpha-\frac{1}{2}\arccos\left(\frac{\theta^2}{8n^2\pi^2}-1\right)\\
\phi_2&=&3\phi_1-2\alpha
\end{eqnarray*}

\noindent
and shows that they all perform with a fidelity $1-{\cal O}(\epsilon^6)$. We 
note that it is only necessary to show the sequence works for one angle, 
say $\alpha=0$, with other axes accounted for by phase-shifting each pulse by 
some angle $\alpha$.

\section*{Other 6$^{th}$ Order Pulse Sequences}

We now look at whether there are other types of pulse sequences similar to 
those above which can achieve the same fidelity. The most general 3-pulse
sequence is,
\begin{eqnarray*}
W^\epsilon(\phi_1,\phi_2,\phi_3)&=&R^\epsilon(\zeta,\phi_1)\,R^\epsilon(\eta,\phi_2)\,R^\epsilon(\gamma,\phi_3).
\end{eqnarray*}

\noindent
As before we define axes $A,B$ and $C$ in the $XY$-plane of the Bloch sphere
by the angles $\phi_1,\phi_2$ and $\alpha$ respectively.
The first condition to satisfy is that the pulse sequence must be symmetric
so that it can be reduced using the symmetric BCH formula, thus
keeping the fidelity for the total pulse sequence 6$^{th}$ order in $\epsilon$. 
This means we need $\zeta=\gamma$ and $\phi_1=\phi_3$. Also to satisfy the 
condition that $W^{\epsilon}(\phi_1,\phi_2)={\mathbb{I}}$, we require, 
\begin{eqnarray}
2\gamma+\eta&=& 4m\pi,m=1,2,...
\end{eqnarray}

\noindent
We are then left with sequences of the form,
\begin{eqnarray*}
W^\epsilon(\phi_1,\phi_2)&=&R^\epsilon(\gamma,\phi_1)\,R^\epsilon(2(2m\pi-\gamma),\phi_2)\,R^\epsilon(\gamma,\phi_1)
\end{eqnarray*} 

\noindent
Finally, we let the first derivative of the total pulse sequence
with respect to $\epsilon$ equal zero at $\epsilon=0$ and obtain the constraint,
\begin{eqnarray*}
&&0=\frac{\theta}{2}C+\gamma A+(2m\pi-\gamma)e^{-i\frac{\gamma}{2}A}B
e^{i\frac{\gamma}{2}A}\\
&\Rightarrow&\sin\gamma\sin(\phi_2-\phi_1)=0,
\end{eqnarray*}

\noindent
giving $\gamma=p\pi, \eta=2q\pi$ with $p+q=2m$ and $p,q=1,2,...$ The most 
general 3-pulse sequences are 
$R^\epsilon(p\pi,\phi_1)\,R^\epsilon(2q\pi,\phi_2)\,R^\epsilon(p\pi,\phi_1)$
with $\phi_1,\phi_2$ determined from,
\begin{eqnarray*} 
\frac{\theta}{2\pi}C+pA+qA^pBA^p&=&0. 
\end{eqnarray*}

\noindent
For even $p$,
\begin{eqnarray*} 
p\sin\phi_1+q\sin\phi_2&=&\frac{-\theta}{2\pi}\sin(\alpha)\\
p\cos\phi_1+q\cos\phi_2&=&\frac{-\theta}{2\pi}\cos(\alpha)
\end{eqnarray*}

\noindent
meaning that either $\phi_1=\alpha-\arcsin(\frac{q}{p}\sin(\phi_2-\alpha))$
or $\phi_2=\alpha+\arcsin(\frac{p}{q}\sin(\alpha-\phi_1))$. Hence
$p/q\leq 1$ since $-1\leq\frac{p}{q}\sin(\alpha-\phi_1)\leq 1\ \forall\alpha$.
Similarily, $q/p\leq 1$, i.e. $q=p$ and $\phi_1+\phi_2=2\alpha$. 
The same is true for odd $p$ except $3\phi_1-\phi_2=2\alpha$.

\noindent
We are left with the general 3-pulse sequences,
\begin{eqnarray*}
W^{\epsilon}_m(\phi_1,\phi_2)&=&R^\epsilon(m\pi,\phi_1)\,R^\epsilon(2m\pi,\phi_2)\,R^\epsilon(m\pi,\phi_1)
\end{eqnarray*}

\begin{figure}[p]
\begin{center}
\setlength{\unitlength}{1cm}
\begin{picture}(6,10)
\put(-2.5,.5){\includegraphics[width=10cm]{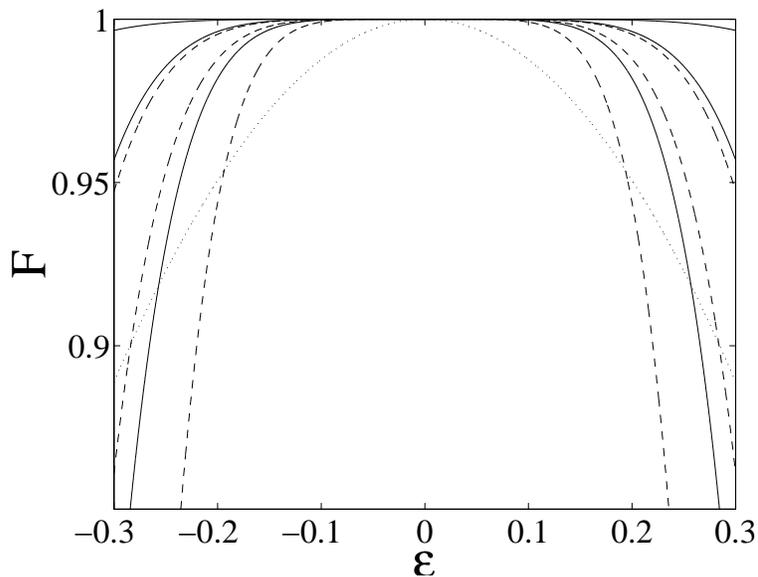}}
\end{picture}
\end{center}
\caption{Fidelity of composite pulse sequences for (a) the 3-pulse sequences 
(solid) and (b) the 5-pulse sequences (dashed), with coefficients for the 
6$^{th}$ order term of the fidelity given in Table 1. The fidelity of the 
single error-prone pulse is also shown (dotted).}
\label{Fig4}
\end{figure}

\noindent
The $Wn$ family is obtained from repeating the $W^{\epsilon}_1(\phi_1,\phi_2)$ 
sequence $n$ times.
When $m=2$, the "passband" Wimperis sequence 
\cite{wimperis,jones_pb} is recovered. Moreover, these are the only 
3-pulse sequences which achieve this 6$^{th}$ order fidelity.
The first three of these composite sequences are plotted in Fig. \ref{Fig4}.
In other 3-pulse sequences, the first order term
in the fidelity always disappears due to the fact that by 
collapsing the entire pulse sequence using the BCH formula to,
\begin{eqnarray*}
BB(\epsilon)&=&e^{-\frac{i\theta(1+\epsilon)}{2}}W^\epsilon(\phi_1,\phi_2)\\
&=&e^{P_1(\epsilon)}
\end{eqnarray*}
 
\noindent
the fidelity is ${\cal F}= 1 + \frac{1}{4}Tr((P_1(\epsilon))^2)+\cdots$ and the
leading term in $P_1(\epsilon)$ is ${\cal O}(\epsilon)$.

So there are no pulse sequences constructed from 3 pulses which 
achieve this 6$^{th}$ order dependence for the fidelity for the resulting 
rotation other than the above family of pulse sequences.
There is, however, the option of creating a 5-pulse sequence by
introducing a third axis. The general form for such a sequence is,
\begin{eqnarray*}
&&R^\epsilon(\zeta,\phi_1)\,
R^\epsilon(\eta,\phi_2)\,
R^\epsilon(\gamma,\phi_3)\,R^\epsilon(\mu,\phi_4)\,R^\epsilon(\nu,\phi_5).
\end{eqnarray*}

\noindent
We define the axes $A,B,C$ and $D$ in the $XY$-plane of the Bloch sphere
by the angles $\phi_1,\phi_2,\phi_3$ and $\alpha$ respectively. 
$D$ is now the axis around which we wish to rotate by $\theta$.
Again, we will require $\zeta=\nu$,$\eta=\mu$ and 
$\phi_1=\phi_5,\phi_2=\phi_4$ in order to keep symmetry in the sequence
and hence the 6$^{th}$ order fidelity dependence. In order that the pulse
sequence is the identity when $\epsilon=0$, we need to satisfy, 
$2\zeta +2\eta+\gamma=4m\pi,\ m=1,2,...$ The first derivative of the total
pulse sequence is zero at $\epsilon=0$ when $\zeta=p\pi$, $\eta=q\pi$ for 
positive integers $p$ and $q$. We arrive at general 5-pulse sequences of the 
form,
\begin{eqnarray*}
&&W^{\epsilon}_{pqr}=\\
&&R^\epsilon(p\pi,\phi_1)\,
R^\epsilon(q\pi,\phi_2)\,
R^\epsilon(2r\pi,\phi_3)\,R^\epsilon(q\pi,\phi_2)\,R^\epsilon(p\pi,\phi_1).
\end{eqnarray*}

\noindent
where $p+q+r=2m$ and $\phi_1$, $\phi_2$ and $\phi_3$ are determined from,
\begin{eqnarray}
\frac{\theta}{2\pi}D+pA+qA^pBA^p+rA^pB^qCB^qA^p&=&0
\end{eqnarray}  

\noindent
One solution to find 5-pulse sequences is at $p=q=r$, analogously to the 3-pulse
sequence case. In this case $p=\frac{2m}{3}$ and, as it must remain an 
integer, $m$ must be a multiple of 3 with now $p=2,4,6...$. Eqn. (11) is
satisfied for $p=2$ when $\phi_1=0$, $\phi_2=\arccos(\frac{\theta-4\pi}{8\pi})$
and $\phi_3=-\phi_2$ for a rotation around the -$X$-axis. However, while the 
fidelity displays a 6$^{th}$ order dependence on the fractional error 
, the coefficient of the leading term, shown in Table \ref{table1},
is so much larger that the sequence is only better than the error-prone pulse 
for small values of $\epsilon$ ($\sim$ 0.2). The situation does not improve
for higher 
values of $p$ and so these sequences are of no real practical use.

Another 5-pulse sequence which achieves the same 6$^{th}$ order dependence
for the fidelity is found by setting $p=1$, $q=2$, $r=1$. Eqn. (11) is now
satisfied when $\phi_1=\arccos(\frac{\theta-4\pi}{4\pi})$, $\phi_2=2\phi_1$, 
$\phi_3=3\phi_1$ for a rotation around the -$X$-axis ($\alpha=\pi$). 
As before, other axes may be accounted for by 
phase-shifting each pulse by the appropriate angle. The fidelity of this
sequence is much better than the previous 5-pulse sequence and 
is quite close to that of the PB1 sequence as seen in Fig. \ref{Fig4}. 
Other sequences can be constructed by varying $p$, $q$ and $r$.
\section*{Conclusion}
We have presented an analysis of the composite pulse sequences
presented by Jones {\it et al.} \cite{JJones,jones_pb} to combat systematic 
pulse-length errors in single-qubit rotations. We have derived an explicit
form for the fidelity and shown how it is possible to set up other 
3-pulse sequences
which achieve the same order error dependence for the fidelity. We have shown 
that there are also  
5-pulse sequences which do achieve the 6$^{th}$ order dependency of the 
fidelity on the error. 

{\bf Acknowledgement}\\
D. McHugh kindly acknowledges support from Enterprise-Ireland Basic Research 
Grant SC/1999/080. The work was also supported by the EC IST FET project QIPDDF-ROSES IST-2001-37150. We thank the referees for their comments.

\begin{table}[H]
\begin{center}
\begin{tabular}{|l|c||l|c|}
\hline
3-pulse&C&5-pulse&C\\
\hline
$W_1^{\epsilon}$(BB1)&4.7&$W_{121}^{\epsilon}$&72.3\\
\hline
$W_2^{\epsilon}$(PB1)&59.1&$W_{112}^{\epsilon}$ &190.6\\
\hline
$W_3^{\epsilon}$&283.4&$W_{222}^{\epsilon}$ & 877.8\\
\hline
\end{tabular}
\caption{The coefficients $C$ in the fidelity expansion $F=1-C\epsilon^6$ 
for six composite pulse sequences which compensate for an 
error-prone $\pi$-pulse around the -$X$-axis.} 
\label{table1}
\end{center}
\end{table}

\end{document}